\documentclass{article}

\usepackage[final]{neurips_2025}

\usepackage[utf8]{inputenc}
\usepackage[T1]{fontenc}
\usepackage{hyperref}
\usepackage{url}
\usepackage{booktabs}
\usepackage{amsfonts}
\usepackage{nicefrac}
\usepackage{microtype}
\usepackage{xcolor}

\usepackage{amsmath,amsfonts}
\usepackage{algorithmic}
\usepackage{array}
\usepackage{textcomp}
\usepackage{stfloats}
\usepackage{url}
\usepackage{verbatim}
\usepackage{graphicx}
\def\BibTeX{{\rm B\kern-.05em{\sc i\kern-.025em b}\kern-.08em
    T\kern-.1667em\lower.7ex\hbox{E}\kern-.125emX}}
\usepackage{balance}

\usepackage{threeparttable}
\usepackage{booktabs}

\usepackage{tikz}
\usepackage{pgfplots}
\usepackage{bbding}
\pgfplotsset{width=8cm,compat=1.9}

\usepackage{amssymb}

\usepackage{float}
\usepackage{graphicx}
\usepackage{subcaption}
\usepackage{caption}

\usepackage{multirow}
\usepackage{multicol}
\usepackage{booktabs}
\usepackage{makecell}
\usepackage{tabularx}
\usepackage{longtable}
\usepackage{threeparttable}
\usepackage{wrapfig}
\usepackage{ulem}

\usepackage{csquotes}

\usepackage{soul}

\usepackage{hyperref}
\hypersetup{hidelinks}
\usepackage[ruled,vlined]{algorithm2e}

\usepackage{enumitem}

\usepackage{balance}

\usepackage{url}
\usepackage{diagbox}
\usepackage{adjustbox}
\usepackage{breqn}

\usepackage{arydshln}
\usepackage{tcolorbox}

\newcommand{\tool}{{E2E-VGuard}\xspace}

\title{\tool: Adversarial Prevention for Production LLM-based End-To-End Speech Synthesis}

\author{%
  Zhisheng Zhang$^{1, 2}$, Derui Wang$^3$, Yifan Mi$^2$, Zhiyong Wu$^{1*}$, 
  \\ \textbf{Jie Gao}$^1$, \textbf{Yuxin Cao}$^5$, \textbf{Kai Ye}$^6$, \textbf{Minhui Xue}$^{3,4}$, \textbf{Jie Hao}$^{2*}$  \\
  $^1$ Shenzhen International Graduate School, Tsinghua University\\
  $^2$ Beijing University of Posts and Telecommunications \quad $^3$ CSIRO’s Data61 \\ 
  $^4$ Responsible AI Research (RAIR) Centre, The University of Adelaide\\
  $^5$ National University of Singapore \quad $^6$ The University of Hong Kong\\
  $^*$ Corresponding authors \\
}

\begin{document}

\maketitle

\begin{abstract}
  Recent advancements in speech synthesis technology have enriched our daily lives, with high-quality and human-like audio widely adopted across real-world applications. However, malicious exploitation like voice-cloning fraud poses severe security risks. Existing defense techniques struggle to address the production large language model (LLM)-based speech synthesis. While previous studies have considered the protection for fine-tuning synthesizers, they assume manually annotated transcripts. Given the labor intensity of manual annotation, end-to-end (E2E) systems leveraging automatic speech recognition (ASR) to generate transcripts are becoming increasingly prevalent, {\it e.g.}, voice cloning via commercial APIs. Therefore, this E2E speech synthesis also requires new security mechanisms.
  To tackle these challenges, we propose \tool, a proactive defense framework for two emerging threats: (1) production LLM-based speech synthesis, and (2) the novel attack arising from ASR-driven E2E scenarios.
  Specifically, we employ the encoder ensemble with a feature extractor to protect timbre, while ASR-targeted adversarial examples disrupt pronunciation. Moreover, we incorporate the psychoacoustic model to ensure perturbative imperceptibility.
  For a comprehensive evaluation, we test 16 open-source synthesizers and 3 commercial APIs across Chinese and English datasets, confirming \tool's effectiveness in timbre and pronunciation protection. Real-world deployment validation is also conducted. Our code and demo page are available at \href{https://wxzyd123.github.io/e2e-vguard/}{https://wxzyd123.github.io/e2e-vguard/}.
\end{abstract}
\section{Introduction}
High-quality synthetic speech based on deepfake techniques~\cite{cosyvoice, GPT-SoVITS} has been applied in daily scenarios, such as video dubbing, and vehicle-mounted voice assistants. Large language models (LLMs)~\cite{gpt2,llama3} have furthered the development of speech synthesis, \textit{i.e.}, text-to-speech (TTS). Current TTS models enhance the synthesis performance by integrating LLMs as a core component for paralanguage features, achieving human-level results. The most advanced technique can be based on an audio foundation model~\cite{step-audio}. TTS models can be divided into two categories, \textit{i.e.}, zero-shot~\cite{cosyvoice} and fine-tuning-based~\cite{vits}. Zero-shot models utilize reference audio as the prompt to clone the voice.
In contrast, fine-tuning-based models require a few minutes of speech samples to replicate the target speaker better. The advancements in speech synthesis, on the one hand, bring huge convenience; on the other hand, they pose a potential security threat in the hands of pirate users. These pirate users may conduct illegal speech synthesis for illegal purposes, such as telecommunication fraud. Therefore, the prevention approach against unauthorized synthesis is of vital importance.

\noindent\textbf{Existing Defenses.} Existing protective methods against voice cloning focus on two main types: (1) Defense based on adversarial examples (AEs), \textit{e.g.}, AntiFake~\cite{antifake}, and AttackVC~\cite{attackvc}. (2) Defense based on unlearnable examples (UEs), \textit{e.g.}, POP~\cite{pop}, and SafeSpeech~\cite{safespeech}. AEs-based protection generates audio AEs through TTS models' encoders to prevent zero-shot voice cloning. In contrast, UEs-based protection utilizes a universal objective to generate model-agnostic perturbation, which disrupts the training phase of TTS models to achieve fine-tuning-based speech synthesis.

\noindent\textbf{Limitations and Challenges.} 
Prior studies have some limitations when considering broader scenarios:
(1) {\it \uline{Industrial-level and LLM-based TTS}}. 
LLMs have advanced the previous deep neural network (DNN)-based speech synthesis. Common approaches employ a speech tokenizer to encode the input waveform into discrete tokens, which are fed into LLMs. The decoded outputs from LLMs then guide the generation module, such as the flow-matching module~\cite{cosyvoice}. 
The key distinction lies in decoding audio signals into discrete tokens rather than continuous embeddings, a direction rarely explored in this protection of LLM-based TTS research. Moreover, voice replication products have emerged in the industry, where voice cloning via API constitutes the focus of this paper.
(2) {\it \uline{End-To-End Scenario}}. The assumption in prior studies is that the text corresponding to the audio has been provided. A more realistic scenario is that, for a customized dataset, the text needs to be obtained by the training party itself. For example, the text transcripts are transcribed through an automatic speech recognition (ASR) system, rather than relying on manually annotated open-source datasets. In fact, commercial APIs typically accept only audio input and rely on an ASR on their backend.

\noindent\textbf{Why is End-to-End Fine-Tuning Important?} 
End-to-end fine-tuning aims to ensure that both input and output data are exclusively of audio type and integrates an ASR system into the training process for text recognition. Based on this, two research questions (RQ) should be considered. \textbf{RQ1:} {\it \uline{Why fine-tuning?}} On the one hand, some models, such as VITS~\cite{vits}, only support fine-tuning without zero-shot capabilities. On the other hand, fine-tuning can achieve better synthetic performance than relying solely on a single sample in zero-shot scenarios. \textbf{RQ2:} {\it \uline{Why end-to-end fine-tuning?}} Adversaries often collect audio data from public social platforms like YouTube and Bilibili, where the audio does not come with corresponding text data. Manually annotating text is typically time-consuming and labor-intensive. Therefore, leveraging an ASR system based on deep learning techniques is a more practical choice due to its efficiency and high recognition accuracy. In the real world, the industrial product for speech synthesis trains a new speaker via an API connection with only audio input. The supplemental ASR system is utilized for automatic recognition in their service.

\noindent\textbf{Our Solution and Contributions.}
To counter the LLM-based and end-to-end scenarios, we propose \emph{\tool}, a proactive defense framework that disrupts both timbre and pronunciation. For the timbre, we introduce untargeted and targeted speaker protection based on the proposed feature loss, which utilizes ensembled encoders and a feature extractor to obtain audio features for LLM-based TTS, resulting in dissimilar synthetic speeches to achieve identification protection. For the pronunciation, \tool generates the audio AEs to fool the ASR system with incorrectly recognized text and disrupt model's learning process of the text and pronunciation. Moreover, to realize imperceptibility, we introduce the psychoacoustic model~\cite{psychoacoustics} to add the perturbation within a specific frequency domain, reducing the detection by the human ears.
For a comprehensive evaluation, we conduct experiments on both English and Chinese datasets, verifying their effectiveness and transferability across 16 open-source, 3 commercial models, and 7 ASR systems. \tool is robust against sophisticated data augmentation and perturbation removal techniques. Moreover, we have validated the \tool's robustness in the real world. 
Our contributions can be summarized as follows:

\begin{itemize}[leftmargin=*]
    \item We introduce a more realistic and challenging scenario of end-to-end fine-tuning-based speech synthesis, and we propose a proactive framework, \tool, to protect individual information.
    \item We consider defensive waveform disruption from the perspectives of timbre and pronunciation. For the timbre disruption, we propose a feature objective based on the encoder ensemble and feature extractor. For the pronunciation disruption, we utilize AEs against ASR systems to fool TTS models with incorrect text and impact pronunciation.
    \item We utilize the psychoacoustic model with $\ell_2$-norm to enhance the perturbation imperceptibility for better human audible perception.
    \item We evaluate the effectiveness, transferability, and robustness of \tool through comprehensive experiments across diverse settings: 19 TTS models (including 16 open-source and 3 commercial), 7 ASR systems, and 3 English and Chinese datasets.
\end{itemize}

\section{Related Work}
\noindent\textbf{Speech Synthesis.} Modern speech synthesis can be primarily categorized into two types: DNN-based and LLM-based architectures. The former mainly focuses on building a synthesizer~\cite{vits,styletts2}, while the latter integrates the LLM with a synthesizer and encodes audio into discretized tokens~\cite{cosyvoice}. The LLM captures the prosody and semantic features, while the synthesizer captures timbre and environmental information~\cite{cosyvoice}. Recently, {\it \uline{audio foundation models}} have advanced rapidly. During the pre-training phase, models typically acquire basic question-answering capabilities. In the post-training phase, they gain downstream tasks, \textit{e.g.}, TTS, through supervised fine-tuning (SFT). This emerging TTS approach based on speech foundation models is also the focus of this paper, \textit{e.g.}, Step-Audio~\cite{step-audio}.

\noindent\textbf{Voice Protection.} Defenses against synthetic speech can be categorized into proactive and passive defenses~\cite{antifake, hiddenspeaker, black_vc}. We primarily focus on proactive defense techniques, which aim to reduce speaker similarity in synthesized audio at the data source. For example, Huang \textit{et al.}~\cite{attackvc} and Yu \textit{et al.}~\cite{antifake} utilized adversarial examples to disrupt voice cloning. Recently, Zhang \textit{et al.}~\cite{pop,safespeech} proposed unlearnable samples to degrade the quality of speech synthesis systems, thereby defending against fine-tuning-based voice synthesis. In addition, we note recent advances in passive defense techniques, such as the robust deepfake audio detection method proposed by Zhang \textit{et al.}~\cite{deepfake_iclr25}, which aims to mitigate the risks posed by synthetic audio. However, existing approaches remain ineffective against more {\it \uline{advanced TTS models and end-to-end scenarios at the data level}}.
\section{\tool Design}\label{section_design}

\subsection{Threat Model}\label{section_design_threat}
In this section, we analyze the necessity of end-to-end fine-tuning through two examples.

\textbf{ByteDance's API.\footnote{\href{https://www.volcengine.com/docs/6561/133350}{https://www.volcengine.com/docs/6561/133350}}} Taking ByteDance's voice cloning product as an example, third-party users upload audio to the company's server infrastructure via the API. First, the input audio is transcribed into text using an ASR system. Then, both the transcribed text and audio information are fed into the TTS model for voice training, enabling synthesis of target text using the trained voice timbre. This commercial API-based speech synthesis approach also fits into the end-to-end scenario.

\textbf{Open-sourced WebUI Operation.} GPT-SoVITS~\cite{GPT-SoVITS} is an open-source zero-shot voice cloning model and supports few-sample fine-tuning to improve voice similarity. The project provides a WebUI-based fine-tuning workflow: the corresponding text is first obtained utilizing an ASR system, {\it i.e.}, Whisper by OpenAI~\cite{whisper}. 
Subsequently, fine-tuning is performed based on the text and audio, which also satisfies the end-to-end scenario requirements.

From these two examples, we can observe that end-to-end speech synthesis is prevalent in real-world applications. Moreover, ASR-based workflows are gradually becoming mainstream, which typically run in model or server backends and are thus not directly accessible to end-users.

\subsection{Problem Formulation}
\begin{figure}[t]
    \centerline{
    \includegraphics[width=0.95\textwidth]{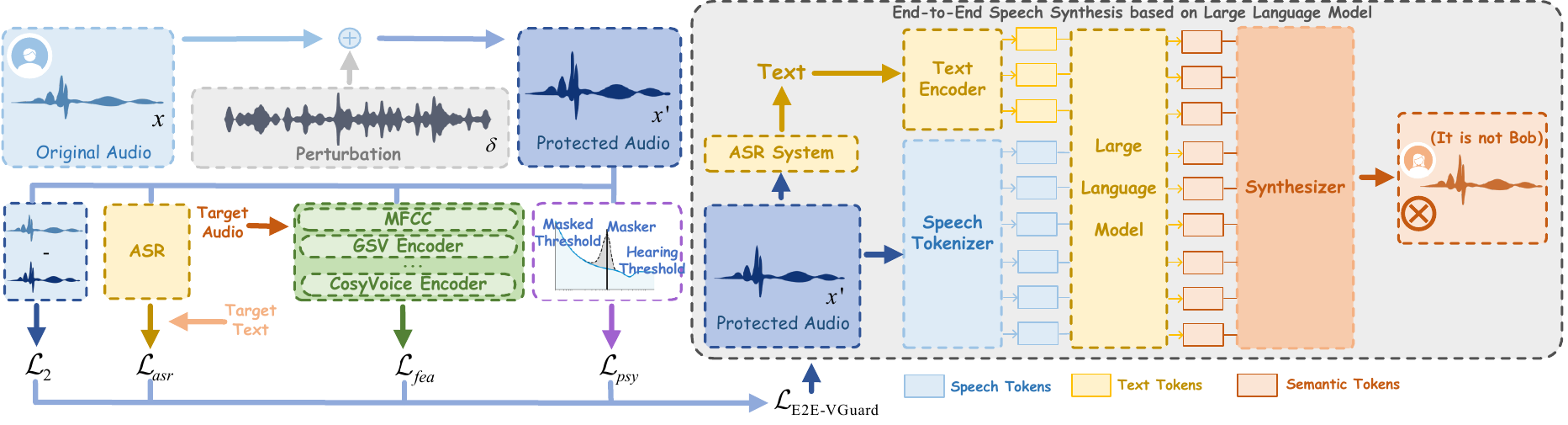}}
    \caption{The workflow of \tool and the end-to-end speech synthesis pipeline with LLM.
    }
    \label{fig_workflow}
\end{figure}

From the perspectives of timbre and pronunciation, we implement \tool for voice anti-cloning. At the timbre level, previous work has focused on targeted~\cite{attackvc, antifake} and untargeted~\cite{pop} attacks, \textit{i.e.}, whether to select a specific target speaker for timbre perturbation. We consider two approaches to timbre protection with a broader defensive selection: targeted and untargeted. For pronunciation protection, adversarial targeted attacks are employed against ASR systems, ensuring that the ASR system transcribes into a specific target text. This is because we aim for the transcribed text to be meaningful rather than gibberish, thereby reducing the adversary's detection of textual alterations. The workflow is shown in Figure \ref{fig_workflow}, and the objective function of \tool can be expressed as:
\begin{equation}
    \begin{aligned}
        & \mathcal{L}(x') = \mathcal{L}_{asr}(x') + \alpha \cdot \mathcal{L}_{fea}(x') + \beta \cdot \mathcal{L}_{psy}(x'), \\
        & \text{s.t.} \quad  ||x' - x||_p \le \epsilon \ \text{and} \ x'\in [-1, 1]^T, \label{eq_total}
    \end{aligned}
\end{equation}

where $\mathcal{L}_{asr}(\cdot)$ represents the loss of the ASR system, $\mathcal{L}_{fea}(\cdot)$ denotes the speech feature loss, and $\mathcal{L}_{psy}(\cdot)$ is the perceptual optimization function for embedded perturbations, \textit{i.e.}, the psychoacoustic model. $\alpha$ and $\beta$ are weight coefficients for multi-task balance. Moreover, $x$ and $x'$ represent the original and protected audio, respectively, with features normalized to fall within the range $[-1, 1]$. The perturbation $\delta:=x'-x$ is bounded by an $\ell_p$-norm constraint $\epsilon$. $T$ is the waveform length. 

\subsection{Timbre Prevention}\label{section_design_timbre}
TTS models typically generate high-quality timbre-similar audio using reference speech samples. We implement timbre-level voice protection to achieve anti-cloning voice protection and ensure cloned audio dissimilarity from the original speakers. Previous research~\cite{antifake, attackvc} primarily employs encoders for timbre extraction and optimization to create timbre divergence between original and target audio. Building upon the timbre encoder ensemble~\cite{antifake}, we further incorporate acoustic features by the MFCC~\cite{mfcc} extractor to better perturb speaker identity features. In designing \tool, we propose two protection methods: targeted and untargeted timbre protection.

\noindent\textbf{Untargeted Timbre Protection.} Untargeted protection maximizes feature distance between original audio $x$ and protected audio $x'$, rendering synthesized audio unrecognizable as the original speaker. Zero-shot TTS models typically employ an encoder to extract cloning-relevant features, \textit{e.g.}, timbre in CosyVoice~\cite{cosyvoice} and style in StyleTTS2~\cite{styletts2}, combining them with pre-trained articulation patterns for speech synthesis. Building on prior findings demonstrating enhanced transferability through encoder ensemble~\cite{antifake, SongBsAb}, we extract timbre and acoustic features through multiple encoders from target TTS models to improve the protective generalizability against unseen models. Moreover, to counter LLM-based TTS, we consider to protect at the audio's original features by changing the discrete tokens obtained by the audio tokenizer for the LLM component with the MFCC extractor to protect articulation and prosodic patterns. The objective function can be formulated as:
\begin{equation}
    \mathcal{L}_{fea}(x') = \sum\limits_{i=1}^k \mathrm{CS}(E_i(x), E_i(x')) + \mathrm{CS}(M(x), M(x')),
\end{equation}
where $k$ is the number of selected encoders, $\mathrm{CS}(\cdot,\cdot)$ is cosine similarity with lower values indicating reduced similarity~\cite{SongBsAb}, $E(\cdot)$ is the timbre encoder, and $M(\cdot)$ denotes the MFCC extractor.

\noindent\textbf{Targeted Timbre Protection.} Targeted protection steers original audio features toward a designated target speaker $x_t$, causing TTS models to synthesize target-like audio. For target selection, we construct a speaker database following AntiFake~\cite{antifake}, choosing the most dissimilar speaker by feature distance for each protected audio. This systematically identifies the most dissimilar speaker in feature space, contrasting with conventional random opposite-gender sampling. The optimization function is:
\begin{equation}
    \mathcal{L}_{fea}(x') = -\,\sum_{i=1}^k \left[\mathrm{CS}(E_i(x_t), E_i(x')) + \mathrm{CS}(M(x_t), M(x')) \right].
\end{equation}
Both methods can effectively protect the speaker's identity. Untargeted protection enables broader adversarial sample exploration due to undefined optimization targets. Targeted protection leverages carefully selected feature-divergent samples for enhanced timbre disruption.

\subsection{Pronunciation Prevention}\label{section_design_psy}
Fine-tuning a TTS model requires pairs of text and audio data to achieve alignment between and pronunciation. For instance, VITS utilizes the monotonic alignment search algorithm to search for the correspondence between time frames and characters. Text data can be obtained through manual annotation or automatic recognition by an ASR system. The former consumes a lot of manpower, time, and costs, therefore, it is more common to employ an ASR system to recognize text information. Previous work~\cite{ZQ-Attack, advddos, weifei_usenix} has shown that ASR systems are relatively vulnerable and susceptible to adversarial examples that interfere with recognition accuracy. Based on this finding, we consider utilizing adversarial examples to disrupt the ASR system's recognition, causing the protected audio to be recognized as a different text. Incorrect text-audio pairs will disrupt the pre-trained model's learning of the pronunciation, preventing the synthesis of audio with the corresponding pronunciation based on the desired text, thereby effectively protecting personal unauthorized audio data.

Adversarial attacks on ASR systems can also be divided into targeted and untargeted types, differing in whether a specific target text is provided when optimizing. The generated text by the untargeted attack is incoherent, while targeted attacks ensure the readability of recognized text by specifying the text, effectively reducing the adversary's awareness of the anomalous recognition text. Therefore, we choose targeted attacks against ASR systems. The selection of the target text affects the effectiveness of adversarial examples. For instance, long audio paired with short target text may result in the latter part of the recognized text retaining the original correct text, necessitating further optimization considerations. Therefore, we need to consider the selected text and its length. In targeted attacks on timbre in Section \ref{section_design_timbre}, since the chosen audio already contains specific pronunciation information different from the original audio, and the application of MFCC also benefits the ASR system's recognition of the target audio text~\cite{ZQ-Attack} in the optimization of adversarial examples, we select the target audio's text as the target text. For untargeted attacks on timbre, we select text of the same length as the audio for different audio, which is more beneficial for adversarial attacks against ASR systems. In summary, the perturbation of pronunciation information can be represented as:
\begin{equation}
    \mathcal{L}_{asr}(x') = \mathcal{F} \left(\mathrm{ASR}(x'), y_t \right),
    \label{eq_asr}
\end{equation}
where $\mathcal{F}$ is the objective function of the ASR system, such as the connectionist temporal classification (CTC)~\cite{ctc} loss for Wav2vec2~\cite{wav2vec2}. $\mathrm{ASR}(\cdot)$ computes the input audio to obtain outputs, such as the probability distribution of recognized words. Additionally, $y_t$ denotes the targeted text.

By optimizing Eq. (\ref{eq_asr}), the adversary utilizes the ASR system and obtains incorrect text, thereby interfering with text-pronunciation alignment, making the synthesized audio unintelligible.

\subsection{Psychoacoustic Model}
The perturbation embedded in a specific region will be masked, making it imperceptible for the human ear to hear the sound in that region, which is known as the \textit{masking effect}. Leveraging this characteristic, we optimize the perception of the embedded perturbation utilizing the psychoacoustic model to enhance the naturalness and imperceptibility. The \textit{masking effect} can be divided into {\it \uline{temporal masking}} and {\it \uline{frequency masking}}. Following the settings of V-Cloak~\cite{v-cloak}, we employ the \textit{frequency masking} part to ensure perturbations are imperceptible to the human ear.

We set the original audio as the \textit{masker} so that the perturbation \textit{(maskee)} remains below the masking threshold. Let $F$ represent the total number of frequencies and $\theta_x$ represent the masking threshold of the original audio $x$, with each element indicating the maximum acceptable perturbation at frequency $f$. Assume $p_x$ represents the log-magnitude power spectral density (PSD) of audio $x$. Therefore, the objective function of the psychoacoustic model can be expressed as:
\begin{equation}
    \mathcal{L}_{psy}(x') = \frac{1}{F}\sum\limits_{f=1}^F \max \left(0, p_{x' - x}(f) - \theta_x(f) \right),
\end{equation}
where $\max(0, \cdot)$ ensures the value is non-negative.

Additionally, we impose constraints on the perturbation through $\ell_2$, as Duan \textit{et al.}~\cite{l2_perception} found that $\ell_2$ performs optimally in correlation with human perception within the $\ell_p$ norm. Therefore, we further reduce the perceptibility of the embedded perturbations by introducing $\mathcal{L}_2$, formulated as:
\begin{equation}
    \mathcal{L}_2(x') = ||x' - x||_2.
    \label{eq_l2}
\end{equation}

To ensure that the protected audio does not exceed the range it should belong to, after obtaining the final protected audio, we map its range back to between -1 and 1 to guarantee it is a normal audio waveform. In conclusion, the algorithm of \tool has been provided in Appendix \ref{section_algorithm}.
\section{Experiments and Analyses}
\textbf{Experiment Organization.} In this section, we introduce our experimental evaluation. 
We first provide our experimental settings in Section \ref{section_exp_settings}. 
Then, we evaluate the effectiveness of end-to-end fine-tuning-based speech synthesis in Section \ref{section_exp_fine-tune}, zero-shot scenarios in Section \ref{section_exp_zero-shot}, and commercial API test in Section \ref{section_exp_api}. Moreover, we explore the effect of each component in Section \ref{section_exp_ablation} and test the robustness of \tool in Section \ref{section_exp_robust}. In the Appendix, there is also an inevitable evaluation. We validate the effect across multilingual and multi-speaker settings in Appendix \ref{section_exp_multi} and various ASR systems in Appendix \ref{section_exp_asr}. Finally, we conduct a human survey of the effectiveness and perception in Appendix \ref{section_exp_human}. 
All of our experiments are conducted on one NVIDIA 4090 GPU. Moreover, the ethical considerations about human study and commercial test are provided in the Appendix, and some limitations and discussions of the \tool have been discussed in Appendix \ref{section_discussion}.

\subsection{Experimental Settings}\label{section_exp_settings}
In this section, we introduce the experimental settings utilized in our experiments.

\noindent\textbf{Synthesizers.}
We select a total of 16 TTS models for evaluation. Section \ref{section_exp_fine-tune} includes 6 models: GPT-SoVITS (GSV)~\cite{GPT-SoVITS}, CosyVoice~\cite{cosyvoice}, Llasa-1B~\cite{llasa}, Llasa-8B~\cite{llasa}, StyleTTS2~\cite{styletts2}, and VITS~\cite{vits}, used for end-to-end fine-tuning tests. Section \ref{section_exp_zero-shot} includes 7 models: Index-TTS~\cite{indextts}, FireRedTTS-1S~\cite{fireredtts-1s}, Step-Audio-TTS~\cite{step-audio}, Spark-TTS~\cite{spark-tts}, XTTS~\cite{xtts}, FishSpeech~\cite{fishspeech}, and Dia-1.6B~\cite{Dia}, for zero-shot validation. Moreover, we test 3 models based on in-context learning rather than speaker encoder for feature extraction in Section \ref{section_exp_zero-shot}: VALLE-X~\cite{vallex}, E2-TTS~\cite{e2-tts}, and F5-TTS~\cite{f5-tts}. Section \ref{section_exp_api} involves 3 commercial APIs: ByteDance, Alibaba, and MiniMax. Notably, Step-Audio-TTS is developed through post-training of a speech foundation model. Further details about open-source models' architectures and sources have been provided in the Appendix \ref{section_tts_models}.

\noindent\textbf{Encoders.} In Section \ref{section_design_timbre}, six encoders serve as feature extractors: posterior encoders from VITS and GSV, MFCC features, WavLM~\cite{wavlm}, CAM++~\cite{cam++} from CosyVoice, and the style encoder from StyleTTS2. Among these, MFCC represents traditional acoustic features, WavLM is a speaker verification system, while the remaining four are timbre or style encoders from TTS systems. This multi-encoder framework improves \tool's cross-model transferability in timbre preservation.

\noindent\textbf{ASR Systems.} Adversaries may employ different ASR systems to recognize text. We conduct model-specific adversarial attacks against ASR systems and can effectively induce misclassification. Seven ASR models are selected, namely Wav2vec2~\cite{wav2vec2}, Whisper (base, small, medium, and large)~\cite{whisper}, Conformer~\cite{conformer}, and CitriNet~\cite{citrinet}. Appendix \ref{section_asr_models} shows different structures in detail.

\noindent\textbf{Datasets.} We selected both single-speaker and multi-speaker datasets in English and Chinese to verify \tool's protection performance across different scenarios, employing LibriTTS~\cite{libritts} for English single-speaker evaluation following \cite{safespeech}, CMU ARCTIC~\cite{cmu_arctic} for English multi-speaker testing, and THCHS30~\cite{thchs30} for Chinese multi-speaker assessment. For each dataset, we have randomly allocated 80\% for training and 20\% for testing. If the model requires a validation set, we utilize 10\% of the training set as the validation set.

\noindent\textbf{Metrics.} The strength and perception metrics are considered.
\begin{itemize}[leftmargin=*]
    \item {\it \uline{Word Error Rate}} (WER)~\cite{pop}. It represents the speech intelligibility. {\it Higher WER reflects lower speech quality}. We utilize a pre-trained ASR model, OpenAI's Whisper with medium size~\cite{whisper}.
    \item {\it \uline{Speaker Similarity}} (SIM)~\cite{safespeech}. SIM measures the speaker similarity of two speeches. {\it Lower SIM reflects lower similarity between original and synthetic speeches}. We employ ECAPA-TDNN~\cite{ecapa_tdnn} to extract speaker embeddings and compute SIM values.
    \item {\it \uline{Signal-to-Noise Ratio}} (SNR)~\cite{safespeech}. SNR reflects the ratio of the embedded perturbation.
    \item  {\it \uline{Perceptual Evaluation of Speech Quality}} (PESQ)~\cite{SongBsAb}: PESQ is an objective perceptual score of the speech quality, ranging from -0.5 to 4.5.
    \item {\it \uline{Mean Opinion Score}} (MOS)~\cite{pop}. MOS is obtained through human interaction as a subjective metric, which measures the human perception of speeches, ranging from 0 to 5.
\end{itemize}

\noindent\textbf{Hyperparameter Settings.} For fine-tuning, we keep the conventional settings with training details in Appendix \ref{section_tts_models}. Moreover, the hyperparameters in Eq. (\ref{eq_total}) are set to balance the effectiveness and imperceptibility of each component. We determine hyperparameters through experiments evaluating both loss values and component effectiveness, ultimately selecting  $\alpha=500$ and $\beta=5\times 10^{-3}$. Additionally, the $\epsilon$ in Eq. (\ref{eq_total}) is $8 / 255$, and we optimize perturbation for 500 iterations.

\subsection{Effectiveness on End-To-End Fine-Tuning Scenarios}\label{section_exp_fine-tune}

To assess the effectiveness and transferability, we utilize \tool to protect the LibriTTS dataset with the untargeted and targeted mode in Section \ref{section_design_timbre} and set the target ASR system as Wav2vec2~\cite{wav2vec2}.

\noindent\textbf{Fine-tuning on Protected Dataset.} After protecting the LibriTTS dataset, users can upload it publicly to social platforms. Adversaries may require these samples unauthorizedly and utilize advanced synthesizers for fine-tuning-based speech synthesis.

\begin{table}[t]
    \caption{Effectiveness and Perception Results of End-To-End Fine-tuning-based Speech Synthesis.  The best and second-best protective results are highlighted with \textbf{bold} and \uline{underlined}, respectively.}
    \resizebox{0.98\linewidth}{!}{
    \scalebox{1.1}{
    \begin{tabular}{ccccccccccccccc}
    \toprule[1pt]\midrule[0.3pt]
    \multirow{2}{*}{\textbf{Method}}
    & \multicolumn{2}{c}{\textbf{GSV}~\cite{GPT-SoVITS}}
    & \multicolumn{2}{c}{\textbf{CosyVoice}~\cite{cosyvoice}}
    & \multicolumn{2}{c}{\textbf{Llasa-1B}~\cite{llasa}}
    & \multicolumn{2}{c}{\textbf{Llasa-8B}~\cite{llasa}}
    & \multicolumn{2}{c}{\textbf{StyleTTS2}~\cite{styletts2}} 
    & \multicolumn{2}{c}{\textbf{VITS}~\cite{vits}}
    & \multicolumn{2}{c}{\textbf{Imperceptibility}} 
    \\
    \cmidrule(r){2-3} \cmidrule(lr){4-5} \cmidrule(lr){6-7} \cmidrule(lr){8-9}
    \cmidrule(lr){10-11} \cmidrule(lr){12-13} \cmidrule(lr){14-15}
    & WER($\uparrow$) & SIM($\downarrow$) 
    & WER($\uparrow$) & SIM($\downarrow$) 
    & WER($\uparrow$) & SIM($\downarrow$) 
    & WER($\uparrow$) & SIM($\downarrow$)
    & WER($\uparrow$) & SIM($\downarrow$) 
    & WER($\uparrow$) & SIM($\downarrow$)
    & SNR($\uparrow$) & PESQ($\uparrow$)\\
    \midrule
    clean
        & 3.434 & 0.685 & 4.288	& 0.700  & 3.157 & 0.643
        & 7.449 & 0.643 & 1.895	& 0.731 & 7.796 & 0.710
        & - & -\\
    AttackVC~\cite{attackvc}
        & 5.205 & 0.636 & 5.531	& 0.688 & 15.201 & 0.569
        & 9.800 & 0.593 & 2.056 & 0.674 & 9.039	& 0.631
        & -2.456 & {\bf 3.890} \\
    AntiFake~\cite{antifake}
        & 28.846 & \uline{0.149} & 7.841 & \uline{0.232} & 22.391 & \uline{0.250}
        & 15.500 & 0.284 & 3.623 & 0.283 & 41.491 & 0.257
        & 12.839 & 1.759\\
    POP~\cite{pop}
        & 3.573 & 0.671 & 4.452 & 0.715 & 7.283 & 0.684
        & 5.692 & 0.675 & 1.756 & 0.743 & 13.281 & 0.685 
        & 18.425 & \uline{3.318} \\
    POP+ESP~\cite{pop}
        & 40.308 & 0.268 & 10.312 & 0.259 & 27.343 & 0.280
        & 34.639 & 0.297 & 7.770 & 0.298 & 55.811 & \uline{0.149}
        & 11.246 & 1.671\\
    SafeSpeech~\cite{safespeech}
        & 44.777 & 0.339 & 8.596 & 0.459 & 9.367 & 0.288 
        & 16.970 & \uline{0.269} & 5.215 & 0.366 & \uline{105.524} & 0.180
        & 7.647 & 1.412 \\
    \midrule
    \textbf{\tool (UT)}
        & \uline{66.471} & {\bf 0.123} & \uline{21.566} & {\bf 0.091} 
        & {\bf 74.956} & {\bf 0.155}
        & \uline{80.221} & {\bf 0.134} & \uline{45.836} & {\bf 0.082} 
        & 95.740 & {\bf 0.106}
        & \uline{18.523} & 1.949\\
    \textbf{\tool (T)}
        & {\bf 94.812} & 0.284 & {\bf 72.143} & 0.375 & \uline{63.945} & 0.442
        & {\bf 89.510} & 0.310 & {\bf 54.732} & \uline{0.229} 
        & {\bf 125.299} & 0.245
        & {\bf 20.470} & 2.324\\
    \midrule[0.3pt]\bottomrule[1pt]
    \end{tabular}
    }
    }
    \label{table_exp_fine-tune}
\end{table}

\noindent\textbf{Speech Synthesis and Evaluation.} TTS models possess the capabilities of speech synthesis after fine-tuning. Fine-tuning-based models can generate speeches with speaker ID and synthesized text, while zero-shot models require reference audio and synthesized text for feature extraction and cloning. Table \ref{table_exp_fine-tune} shows the experimental results across different TTS models after fine-tuning on datasets protected by different strategies. We can find that our protected \tool can achieve an outstanding protective strength than baselines in terms of timbre (SIM) and pronunciation (WER). For fine-tuning-based models, the \tool achieves an average increase of 19.775\% (targeted, T) in WER compared to the best baseline values while reducing SIM by an average of 0.043 (UT), indicating lower speech intelligibility and similarity. The protection effect improves more significantly on zero-shot models, with WER increasing by an average of 32.841\% (UT) and 50.060\% (T) and SIM decreasing by 0.119 (UT). It demonstrates that audio protected by the \tool effectively safeguards private information, prevents high-quality speech synthesis, and exhibits strong transferability across models.

\noindent\textbf{Perception Analyses.} The embedded perturbation should not interfere with the normal utilization of speeches. Table \ref{table_exp_fine-tune} presents the simulated perception metrics of different baselines. The SNR values are higher than all baselines, representing the lowest noise ratio and quality disruption.

\subsection{Effectiveness on End-To-End Zero-shot Scenarios}\label{section_exp_zero-shot}

In this section, we conduct zero-shot end-to-end speech synthesis on seven industrial-level and LLM-based TTS models. The reference audio transcripts are automatically obtained through an ASR system. Table \ref{table_exp_zero-shot} presents the test results, where our \tool (UT) achieves SOTA performance in voice timbre preservation across all models than baselines. Regarding pronunciation prevention, the average WER values of \tool are 21.603\% for UT and 23.604\% for T, respectively, outperforming AntiFake's 4.932\% and SafeSpeech's 19.255\%. This indicates the synthesized audio demonstrates both dissimilar timbre characteristics and reduced pronunciation clarity.

\begin{table}[t]
    \caption{Protective performance across industrial-level and LLM-based models of \tool under zero-shot end-to-end speech synthesis.}
    \resizebox{0.98\linewidth}{!}{
    \scalebox{1.1}{
    \begin{tabular}{ccccccccccccccc}
    \toprule[1pt]\midrule[0.3pt]
    \multirow{2}{*}{\textbf{Method}}
    & \multicolumn{2}{c}{\textbf{Index-TTS}~\cite{indextts}}
    & \multicolumn{2}{c}{\textbf{FireRedTTS-1S}~\cite{fireredtts-1s}}
    & \multicolumn{2}{c}{\textbf{Step-Audio-TTS}~\cite{step-audio}}
    & \multicolumn{2}{c}{\textbf{Spark-TTS}~\cite{spark-tts}}
    & \multicolumn{2}{c}{\textbf{XTTS-v2}~\cite{xtts}} 
    & \multicolumn{2}{c}{\textbf{FishSpeech}~\cite{fishspeech}} 
    & \multicolumn{2}{c}{\textbf{Dia-1.6B}~\cite{Dia}} 
     \\
    \cmidrule(r){2-3} \cmidrule(lr){4-5} \cmidrule(lr){6-7} \cmidrule(lr){8-9}
    \cmidrule(lr){10-11} \cmidrule(lr){12-13} \cmidrule(lr){14-15}
    & WER($\uparrow$) & SIM($\downarrow$) 
    & WER($\uparrow$) & SIM($\downarrow$) 
    & WER($\uparrow$) & SIM($\downarrow$) 
    & WER($\uparrow$) & SIM($\downarrow$)
    & WER($\uparrow$) & SIM($\downarrow$) 
    & WER($\uparrow$) & SIM($\downarrow$)
    & WER($\uparrow$) & SIM($\downarrow$)\\
    \midrule
    clean
        & 3.547 & 0.674 & 1.382 & 0.655
        & 2.508 & 0.579 & 1.341 & 0.666
        & 1.258 & 0.555 & 1.848 & 0.497
        & 4.526 & 0.581\\
    AntiFake~\cite{antifake}
        & 3.685 & {\bf 0.147} & 3.152 & \uline{0.276} 
        & 2.395 & 0.206 & 4.214 & 0.243
        & \uline{7.115} & \uline{0.193} & 8.495 & 0.178
        & 5.471 & 0.283\\
    POP+ESP~\cite{pop}
        & 3.730 & 0.282 & 4.491 & 0.366
        & \uline{5.845} & 0.312 & 30.667 & 0.180
        & 7.059 & 0.358 & 6.998 & \uline{0.129}
        & 6.266 & 0.315 \\
    SafeSpeech~\cite{safespeech}
        & {\bf 5.532} & 0.244 & 6.098 & 0.339
        & 5.764 & 0.334 & 23.866 & \uline{0.144}
        & {\bf 13.522} & 0.230 & {\bf 21.335} & 0.197
        & \uline{58.669} & 0.250 \\
    \midrule
    \textbf{\tool (UT)}
        & \uline{4.474} & \uline{0.196} & \uline{6.528} & {\bf 0.226}
        & {\bf 8.156} & {\bf 0.008} & \uline{33.357} & {\bf 0.174}
        & 5.761 & {\bf 0.173} & \uline{9.746} & {\bf 0.127}
        & {\bf 83.200} & {\bf 0.208} \\
    \textbf{\tool (T)}
        & 2.667 & 0.441 & {\bf 40.338} & 0.367
        & 3.245 & \uline{0.128} & {\bf 72.522} & 0.260
        & 3.014 & 0.455 & 7.633 & 0.218
        & 35.811 & \uline{0.248} \\
    \midrule[0.3pt]\bottomrule[1pt]
    \end{tabular}
    }
    }
    \label{table_exp_zero-shot}
\end{table}

The experimental results from Section \ref{section_exp_fine-tune} and Section \ref{section_exp_zero-shot} demonstrate that our proposed \tool effectively protects the latest open-source industrial-level and LLM-based models from timbre and pronunciation perspectives. The method achieves current SOTA performance levels in safeguarding personal information security, regardless of whether fine-tuning-based or zero-shot speech synthesis.

\begin{table}[tbp]
\centering
\begin{minipage}{0.48\textwidth}
    \centering
    \caption{Evaluation on ICL-based TTS models.}
    \resizebox{0.9\linewidth}{!}{
    \begin{threeparttable}
    \begin{tabular}{cccccccc}
    \toprule[1pt]\midrule[0.3pt]
    \multirow{2}{*}{\textbf{Method}} 
    & \multicolumn{2}{c}{\textbf{F5-TTS}~\cite{f5-tts}}
    & \multicolumn{2}{c}{\textbf{E2-TTS}~\cite{e2-tts}}
    & \multicolumn{2}{c}{\textbf{VALLE-X}~\cite{vallex}} \\
    \cmidrule(r){2-3}\cmidrule(r){4-5}\cmidrule(r){6-7}
    & WER($\uparrow$) & SIM($\downarrow$) 
    & WER($\uparrow$) & SIM($\downarrow$)
    & WER($\uparrow$) & SIM($\downarrow$) \\
    \midrule
     clean & 
        4.268 & 0.676 & 5.401 & 0.678 & 14.450 & 0.519\\
    AntiFake~\cite{antifake} & 
        4.303 & \uline{0.282} & 4.004 & \uline{0.269} & 96.469 & 0.249 \\
    
    \midrule
    \textbf{\tool (UT)} &
        \uline{10.776} & {\bf 0.053} & \uline{7.064} & {\bf 0.138}
        & {\bf 129.483} & {\bf 0.175} \\
    \textbf{\tool (T)} & 
        {\bf 70.034} & 0.319 & {\bf 84.913} & 0.372 
        & \uline{88.707} & \uline{0.176} \\

    \midrule[0.3pt]\bottomrule[1pt]
    \end{tabular}
    \end{threeparttable}
    }
    \label{table_exp_icl}
\end{minipage}
\hfill
\begin{minipage}{0.48\textwidth}
    \centering
    \caption{The ablation study of the \tool.}
    \resizebox{0.9\linewidth}{!}{
    \begin{threeparttable}
    \begin{tabular}{ccccccc}
    \toprule[1pt]\midrule[0.3pt]
    \multirow{3}{*}{\textbf{w/ o}} 
    & \multicolumn{4}{c}{\textbf{Effectiveness}}
    & \multicolumn{2}{c}{\multirow{2}{*}{\textbf{Imperceptibility}}} \\
    \cmidrule(r){2-5}
    & \multicolumn{2}{c}{\textbf{VITS}}
    & \multicolumn{2}{c}{\textbf{GSV}} \\
    \cmidrule(r){2-3}\cmidrule(r){4-5}\cmidrule(r){6-7}
    & WER($\uparrow$) & SIM($\downarrow$)
    & WER($\uparrow$) & SIM($\downarrow$)
    & SNR($\uparrow$) & PESQ($\uparrow$) \\
    \midrule
    $\mathcal{L}_{psy}$ \& $\mathcal{L}_{2}$
        & {\bf 119.242} & {\bf 0.101}
        & {\bf 90.436} & {\bf 0.081}
        & 12.942 & 1.544 \\
    $\mathcal{L}_{fea}$
        & \uline{101.407} & 0.177
        & 65.446 & 0.409
        & \uline{15.065} & \uline{1.811} \\
    $\mathcal{L}_{asr}$
        & 94.940 & \uline{0.102}
        & 49.059 & 0.124
        & 13.724 & 1.572\\
    \midrule
    \textbf{\tool (UT)}
        & 95.740 & 0.106
        & \uline{66.471} & \uline{0.123}
        & {\bf 18.523} & {\bf 1.949} \\
    \midrule[0.3pt]\bottomrule[1pt]
    \end{tabular}
    \end{threeparttable}
    }
    \label{table_exp_ablation_1}
\end{minipage}
\end{table}

\noindent\textbf{Evaluation on ICL-based TTS models.} Many existing TTS models extract timbre representations of target speakers through a speaker encoder, and \tool also utilizes this approach by integrating multiple speaker encoders to achieve timbre-level prevention. Additionally, some models obtain information such as timbre features through in-context learning (ICL) rather than using a speaker encoder.
Table \ref{table_exp_icl} presents our experimental results on three ICL-based models. The results demonstrate that \tool maintains SOTA voice protection performance on ICL-based models and exhibits strong transferability. This effectiveness stems from our speaker encoder ensemble technique, which successfully hides or modifies the timbre information of the original speaker. Consequently, the timbre prevention of the proposed E2E-VGuard does not rely on the specific speaker encoder.

\subsection{Evaluation via Commercial APIs}\label{section_exp_api}
\begin{figure*}[t]
    \centering
    \begin{subfigure}{0.3\textwidth}
        \centering
        \includegraphics[width=\textwidth]{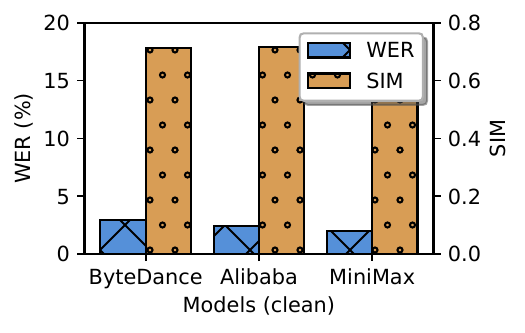} 
        \caption{Clean mode.}
        \label{fig_api_clean}
    \end{subfigure}
    \begin{subfigure}{0.3\textwidth}
        \centering
        \includegraphics[width=\textwidth]{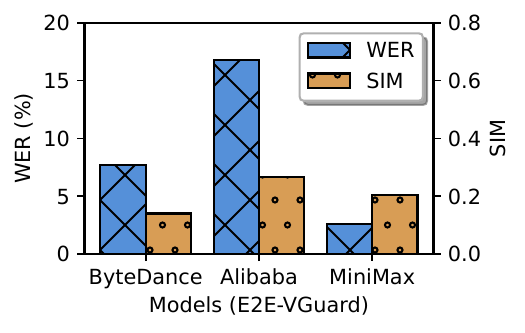}
        \caption{\tool mode.}
        \label{fig_api_p}
    \end{subfigure}
    \caption{Evaluation of protective performance on commercial APIs.}
    \label{fig_api}
\end{figure*}
Voice cloning through commercial APIs is relatively convenient, requiring only audio data input to train the target speaker. Once the server generates the trained speaker ID, speech synthesis can be readily implemented. This approach eliminates the need for local model deployment while achieving high-quality synthesis, as the underlying models are black-box systems that demand robust defense mechanisms. We select three common commercial products supporting voice replication, including ByteDance, Alibaba, and MiniMax (represented by company names), for evaluation.

As shown in Figure \ref{fig_api}, our experimental results reveal that compared with unprotected audio, \tool reduces the average SIM score from 0.689 to 0.203 while increasing WER values. This demonstrates \tool's strong transferability, effectively safeguarding voiceprint information even in black-box scenarios involving commercial APIs. These performance improvements across similarity and pronunciation metrics confirm its practical effectiveness for real-world deployment.

\subsection{Ablation Study}\label{section_exp_ablation}
In the ablation study, we explore the functionality of our proposed optimization objectives and the hyperparameter selection for the trade-off of the perturbative performance and imperceptibility.

\noindent\textbf{Component Analyses.} In this part, we explore the role of each component in Eq. (\ref{eq_total}).
We separately investigate the impacts of $\mathcal{L}_{asr}$, $\mathcal{L}_{fea}$, and $\mathcal{L}_{psy}$ \& $\mathcal{L}_2$ by removing each component from \tool and evaluating the resulting protection effectiveness. Table \ref{table_exp_ablation_1} presents the results of this ablation study. We observe that removing the perceptual optimization module, \textit{i.e.}, $\mathcal{L}_{psy}$ \& $\mathcal{L}_2$, achieves better protection but significantly increases perceptual disruption to the original audio due to the lack of perceptual alignment of noise, resulting in a low SNR of 12.942. To directly examine the effects of $\mathcal{L}_{asr}$ and $\mathcal{L}_{fea}$, we optimize each term individually. When $\mathcal{L}_{fea}$ is removed and only $\mathcal{L}_{asr}$ is retained, the SIM value on the GSV model is relatively high at 0.409, indicating significant leakage of voiceprint information. Conversely, when $\mathcal{L}_{asr}$ is removed and only $\mathcal{L}_{fea}$ is retained, the disruption of text-pronunciation alignment diminishes, with the WER on the GSV model dropping to 49.059\%. When all three components are present, they collectively provide effective protection at both the pronunciation and timbre levels while maintaining better imperceptibility of perturbations.

\noindent\textbf{Hyperparameter Selection.} In the Eq. (\ref{eq_total}), we employ hyperparameters $\alpha$ and $\beta$ to balance the effectiveness of \tool's protection and the imperceptibility of the embedded, with values empirically set to 500 and $5\times 10^{-3}$~\cite{v-cloak}, respectively. The value of $\alpha = 500$ is chosen to amplify the loss for speaker identity protection, aligning its optimization scale approximately with that of $\mathcal{L}_{\text{asr}}$. $\beta$ is used to trade off the protection effectiveness and the perception quality of the perturbation. Specifically, a larger $\beta$ (such as $5\times 10^{-2}$) yields better perception quality but weaker protection, \textit{e.g.}, the speaker similarity degrading, whereas a smaller $\beta$ (such as $5 \times 10^{-4}$) enhances protection at the cost of reduced perception quality \textit{e.g.}, SNR lower than 15. Through empirical evaluation with various $\beta$ values, we find that $\beta = 5 \times 10^{-3}$ satisfies an approximate trade-off between protection and quality of perception.

\subsection{Robustness Test}\label{section_exp_robust}
In real-world scenarios, strong adversaries can find the obtained dataset with specific modifications, and adversarial techniques may be employed to improve synthesis quality. In this part, following~\cite{pop,safespeech}, we conduct the robustness test against perturbation removal and advanced data augmentation techniques. Moreover, we evaluate the effectiveness and robustness of \tool in the real world. 

\noindent\textbf{Perturbation Removal.} High-quality speech synthesis often requires high-quality input audio without audible perturbation~\cite{safespeech}. Therefore, adversaries may utilize perturbation removal techniques to improve the quality of training samples and weaken unknown strategies users adopt before uploading. We refer to the use of two efficient denoising techniques~\cite{safespeech}, spectral gating (SG), and a DNN-based model named \textit{denoiser}~\cite{denoiser} to denoise each protected audio sample. This experiment is conducted on two models, \textit{i.e.}, VITS and GSV. For the SG method, the WER and SIM on the VITS model are 51.005\% and 0.224, respectively, while on the GSV model, the WER and SIM are 31.958\% and 0.251, respectively. The \textit{denoiser} can nearly remove the audible background noise. We test the protective performance utilizing the \textit{denoiser} for denoising. On the GSV model, the WER and SIM are 23.10\% and 0.243, respectively. The WER and SIM on the VITS model are 34.10\% and 0.261, respectively. 
This shows that even after removing the audible noise, \tool can still protect speaker privacy, especially at the level of the speaker identity, with an average SIM value of only 0.238 and 0.252 across these two models using the SG and \textit{denoiser}, respectively. The WER on the VITS model exceeds 50\% after denoising using the SG method, effectively disrupting pronunciation.

\noindent\textbf{Data Augmentation.} Data augmentation is used to alter the specific structures of embedded perturbations, thereby reducing effects. We consider three categories of data augmentation techniques:

\begin{itemize}[leftmargin=1em, topsep=0pt, itemsep=0pt, parsep=0pt]
    \item \textbf{Adversarial Defender}~\cite{waveguard}: Hussain \textit{et al.}~\cite{waveguard} discovered that in the audio field, certain adversarial defense methods can effectively disrupt adversarial audio examples, {\it e.g.}, proposed \tool. These adversarial techniques include: Down-sampling and Up-sampling (RS), Mel-spectrogram Extraction and Inversion (Mel), Quantization-Dequantization (Q-D), and Filtering.
    \item \textbf{Audio Processor}~\cite{pop}: We consider speech-processing techniques to simulate real-world operations, following Zhang \textit{et al.}~\cite{pop}, including Speed Adjustment (Speed), adding Gaussian noise (Gaussian), Time Masking (TiM), Pitch Shifting (PS), MP3 compression, and Tanh Distortion (Tanh).
    \item \textbf{Filters}~\cite{pop}: Filtering techniques are commonly used to alter perturbations. We consider three types of filter techniques: Band-Pass Filter (BPF), Low-Pass Filter (LPF), and High-Pass Filter (HPF).
\end{itemize}

Table \ref{table_augmentation} shows the results of data augmentations. We observe that the Mel technique significantly improves the intelligibility of synthesized audio, with WER values decreasing by 39.194\% and 19.825\% on the VITS and GSV models, respectively. However, the SIM values remain high. Although data augmentation can disrupt the perturbation and reduce its protective effect, the embedded perturbation persists, and transforming the audio inherently degrades its quality, \textit{e.g.}, Mel, and Filters.

\begin{table}[t]
    \centering
    \caption{Results under data augmentation and defensive methods. The \underline{underline} indicates the most significant decreases in protection compared to training without augmentation (``w/ o'' in the Table) .}
    \resizebox{0.98\linewidth}{!}{
    \begin{tabular}{ccccccccccccccccc}
    \toprule[1pt]\midrule[0.3pt]
        \multirow{2}{*}{\textbf{Model}} 
        & \multirow{2}{*}{\textbf{Metric}}
        & \multirow{2}{*}{\textbf{w/ o}} 
        & \multicolumn{4}{c}{\textbf{Adversarial Defender}~\cite{waveguard}}
        & \multicolumn{6}{c}{\textbf{Audio Processor}~\cite{pop}} 
        & \multicolumn{3}{c}{\textbf{Filters}~\cite{pop}} \\
        \cmidrule(r){4-7} \cmidrule(lr){8-13} \cmidrule(r){14-16}
        & & & Resample & Mel & Q-D 
        & Filtering & Speed & Gaussian
        & TiM & PS & MP3
        & Tanh & BPF & LPF & HPF \\
        \midrule
        \multirow{2}{*}{VITS}
            & WER($\uparrow$)   & 96.735 & 94.827 & {\bf 55.633} & 103.247 & 93.552 
                    & 84.101 & 84.881 & 95.654 & 91.796 & \uline{82.398}
                    & 95.441 & 97.258 & 95.553 & 136.552 \\
            & SIM($\downarrow$)   & 0.113 & 0.115 & 0.122 & 0.082 & 0.128 
                    & 0.080 & {\bf 0.163} & 0.098 & 0.099 & 0.125
                    & \uline{0.143} & 0.136 & 0.118 & 0.045 \\
        \midrule
        \multirow{2}{*}{GSV}
            & WER($\uparrow$)   & 69.148 & 55.035 & {\bf 35.210} & 75.011 & 57.497 
                    & 77.642 & 44.232 & 77.165 & 68.607 & \uline{41.903}
                    & 71.446 & 83.050 & 52.882 & 48.036 \\
            & SIM($\downarrow$)   & 0.074 & \uline{0.195} & 0.158 & 0.117 & 0.147 
                    & 0.044 & 0.128 & 0.109 & 0.049 & 0.154
                    & 0.126 & -0.046 & 0.135 & {\bf 0.229} \\
    \midrule[0.3pt]\toprule[1pt]
    \end{tabular}
    }
    \label{table_augmentation}
  \end{table}

\begin{figure*}[t]
    \centering
    \begin{subfigure}{0.3\textwidth}
        \centering
        \includegraphics[width=\textwidth]{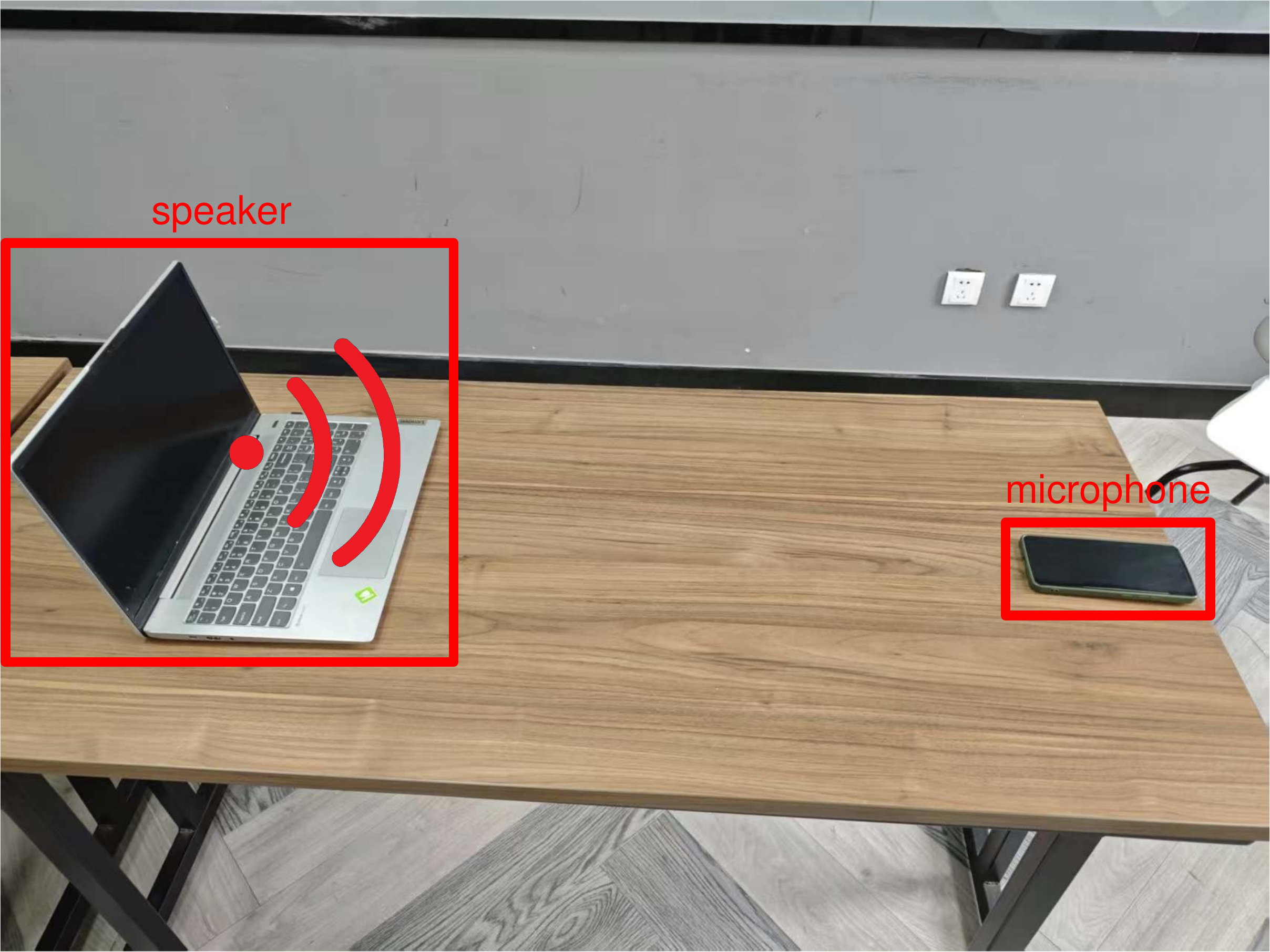} 
        \caption{Real-world environment.}
        \label{fig_ota_env}
    \end{subfigure}
    \begin{subfigure}{0.3\textwidth}
        \centering
        \includegraphics[width=\textwidth]{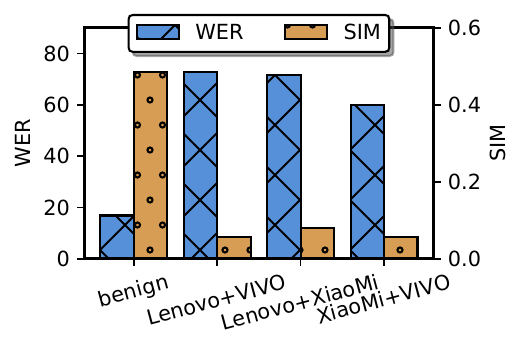}
        \caption{Results.}
        \label{fig_ota_results}
    \end{subfigure}
    \caption{Robust test in the real world. (a) shows the experimental environment. (b) The experimental results. \enquote{Lenovo+VIVO} represents the \enquote{speaker-microphone}.}
    \label{fig_ota}
\end{figure*}

\noindent\textbf{Real-World Robustness.} After users upload audio protected by \tool, adversaries may utilize various types of microphones to record and collect the played audio in the real world for voice cloning. To verify the robustness of \tool in over-the-air scenarios, we employed different speakers to play the audio as the speaker and different microphones to record the audio as adversaries. We conduct experiments in a quiet environment with background noise averaging 22 dBA (measured by taking the average over 10 seconds). We apply the built-in speaker of a Lenovo Laptop to play the audio and record the audio using VIVO and XiaoMi phones placed approximately one meter away from the speaker to simulate the adversaries as shown in Figure \ref{fig_ota_env}. In each test, we play 10 audio samples and ensure that the speaker's loudness averages 46 dBA. Finally, for the recorded audio, we employ the GSV model for fine-tuning and cloning due to its excellent few-shot cloning capability~\cite{GPT-SoVITS}. The original synthesized results without perturbation after recording are 16.837\% and 0.485 of WER and SIM values with high speaker consistency. The average WER and SIM are 72.615\% and 0.068, respectively, in Figure \ref{fig_ota_results}, indicating excellent protection of voice timbre and pronunciation in real-world scenarios. Additionally, when we replace the speaker with a mobile device, a XiaoMi phone, and use the VIVO phone as the recording device, the protective effect remains high. This experiment demonstrates the robustness of \tool in the real world, as the over-the-air transmission acts as a form of data augmentation~\cite{SongBsAb}, and Section \ref{section_exp_robust} illustrates the effectiveness of \tool in handling data augmentation.
\section{Conclusion}
This paper focuses on the current mainstream industrial-level and LLM-based TTS models. Considering the more practical scenario of end-to-end speech synthesis, we propose a protection technique, \tool, that effectively safeguards audio content from both timbre and pronunciation perspectives. We conduct extensive experiments on various speech synthesis models and multilingual datasets for evaluation. 
Limitations and future work, \textit{e.g.}, time efficiency and the \tool's reliance, are discussed in the appendix.

\section*{Acknowledgment}
We sincerely appreciate anonymous reviewers for their insightful and valuable feedback. 
Zhisheng Zhang, Yifan Mi, and Jie Hao are supported in part by the National Natural Science Foundation of China under Grant No. U21B2020 and the Fundamental Research Funds for the Central Universities under Grant No. 2024ZCJH05. Zhisheng Zhang, Jie Gao, and Zhiyong Wu are supported in part by the National Natural Science Foundation of China under Grant No. 62076144 and the Shenzhen Science and Technology Program under Grant No. JCYJ20220818101014030.

\normalem
\bibliographystyle{unsrt}
\bibliography{references}

\clearpage
\appendix

\section{Discussions and Limitations}\label{section_discussion}

\noindent\textbf{Ethical Considerations.} To verify the subjective perception of \tool in the protected and synthesized audio for human ears, we conduct human subjective testing experiments in Appendix \ref{section_exp_human}. These experiments have received approval from the local Human Ethics Research Committee. In the questionnaire, all recruited volunteers are anonymous and consented to their answers being used only for academic research. We do not collect any information beyond the content of the questionnaire, and we maintain strict confidentiality regarding the responses. All synthesized audio is uniformly discarded after the completion of the subjective survey to ensure no security risks through leakage. Additionally, the experiments described in Section \ref{section_exp_api} involve testing with commercial APIs. Before using the company's products, we have completed real-name authentication and documented the process. Our experimental testing is conducted internally and will not affect the company's operations or generate unintended usage impacts.

\noindent\textbf{Broader Impacts.} This paper primarily focuses on proposing a proactive defense technique. Our intention is positive, aiming to protect individuals' voices from infringement. We will open-source the \tool, grant users the right to use it, and sign relevant disclaimer clauses to ensure that their usage behaviors are not related to the designer and publisher of the \tool. According to the licensing agreement, this will not affect the normal and positive usage of speech synthesis technology. Additionally, our testing using commercial APIs is conducted solely for local user testing and will not impact any company's products or services. This paper will not result in any negative societal impacts.

\noindent\textbf{Subjective Bias.} The conclusions of the subjective experiments are derived from human responses, which can be influenced by the answering circumstances, potentially leading to subjective bias. To reduce the bias in subjective experiments, we calculate the 95\% confidence interval of the MOS values following~\cite{pop} and recruit a sufficient number of volunteers to enhance the reliability of the conclusions. We also implement certain filtering measures to eliminate low-quality responses. According to the conclusions of the subjective experiments, human subjective perception is generally aligned with objective evaluations, indicating that we have minimized the interference of subjective bias on the experimental results.

\noindent\textbf{Limitations of the Target ASR System.} In designing \tool, we focus on protecting against a specific ASR system, aiming to cause errors in the text recognized by the target ASR. This breaks the alignment between text and pronunciation in the TTS model. The reason we do not pursue a universal approach is that previous methods~\cite{ZQ-Attack} based on universal attacks typically rely on clustering multiple ASRs, which consumes more time and computing resources. We aim to simplify our system as much as possible to improve the efficiency of the protection. Therefore, we conduct targeted protection using the targeted ASR and then transfer the results to other models. The experiments in Appendix \ref{section_exp_asr} also demonstrate that our method remains effective when employing different targeted ASR systems.

\noindent\textbf{Time Overhead and Acceleration Strategies.} As the defender, we consider scenarios where users upload audio to the internet after \tool protects audio samples. We test the average time for \tool to protect audio on the LibriTTS dataset using a device equipped with one NVIDIA 4090 GPU with 24 GB of memory. On average, \tool takes 97.982 seconds and 111.495 seconds to protect audio in untargeted and targeted settings, respectively. This protection time is on the same level as the baselines, which are 44.871 seconds for AttackVC~\cite{attackvc} and 203.248 seconds for AntiFake~\cite{antifake}. The shorter optimization time may allow users to protect the target audio more quickly. The additional time overhead for targeted protection with \tool, compared to untargeted protection, mainly comes from selecting a target speaker from the speaker database. Moreover, some acceleration methods can reduce the time overhead, \textit{e.g.}, data batching and multi-GPU parallelization.

\noindent\textbf{Adaptive Adversaries.} In the Section \ref{section_exp_robust}, we evaluate the robustness of \tool from three aspects. The experiments demonstrate that \tool can resist denoising techniques and remains effective against the adversarial example defense techniques proposed by Hussain \textit{et al.}~\cite{waveguard} in the audio domain. Additionally, we test various audio compression and filtering techniques and simulate the adversary's acquisition of audio using different speakers and microphones in real-world scenarios. The \tool can still effectively protect the audio. The reason for \tool's strong robustness lies in its application of various feature encoders to capture information from the latent space of the audio, enabling perturbations to be better embedded and thus resistant to being disrupted by denoising techniques and others.

\noindent\textbf{Encoder Ensemble.} TTS models, especially zero-shot ones, typically design a speaker encoder to extract the timbre embedding of the reference audio. For a specific TTS model, one can perform an adversarial attack on its speaker encoder to mislead the extraction of the target timbre, thereby protecting the original speaker's timbre. In practice, as defenders, we cannot know what type of TTS model the adversary might employ, so our designed proactive defense framework, \tool, should possess transferability. Previous research has shown that clustering encoders from different models can achieve outstanding transferability. Based on this, we utilize an encoder ensemble approach and combine it with a feature encoder to better extract and protect the reference speaker's timbre. The encoders we selected are highly representative and can cover mainstream generative architectures and backbones, such as VAE from VITS~\cite{vits}, diffusion model from StyleTTS2~\cite{styletts2}, and flow matching from CosyVoice~\cite{cosyvoice}.

\noindent\textbf{Eliminating ASR System.} In the scenario described in this paper, we have developed an end-to-end fine-tuning method based on an ASR system that does not require manually labeled text. We propose \tool for timbre and pronunciation protection. However, assuming the adversary has enough human resources to obtain text through manual labeling, the effectiveness of \tool remains a concern. We conduct experiments on the GSV model, using both untargeted and targeted audio protection. For the reference text in fine-tuning, we provide the correct text instead of the text obtained through ASR transcription. The experimental results show that the WER and SIM are 39.659\%, 0.161 (T) and 73.784\%, 0.278 (UT). This indicates that using clean text can still achieve effective protection at the timbre level, and pronunciation will continue to be affected. This demonstrates that the perturbations added by \tool can interfere with the TTS model's learning of pronunciation information. Therefore, \tool can still provide some protective effect even when manually labeled correct text is used for fine-tuning.

\noindent\textbf{``Imperceptibility'' consideration.} In the scenario of our paper, the embedded perturbations should be ``harmless'' to the original audio, meaning that the original text content remains unaltered and the normal usability of the protected audio is unaffected. ``Usability'' represents whether the audio can be utilized normally in our daily lives. Rather than requiring perceptual indistinguishability between the protected and original audio. We have verified through both objective (Section \ref{section_exp_fine-tune}) and subjective experiments (Appendix \ref{section_exp_human}) that the perturbations we generated do not cause huge disruptions to the original audio.
Moreover, from the robustness perspective, assuming strong adversaries can distinguish embedded perturbations, they can utilize adversarial techniques to improve the performance of the synthesized speech, causing privacy leakage. However, the robustness validated in Section 4.6 ensures that the adversary cannot effectively remove the embedded perturbation, thereby enhancing protection efficacy against speech synthesis. Therefore, even if the adversary perceives the perturbations, the robustness of \tool ensures that privacy data is not completely leaked. 

\section{Algorithm}\label{section_algorithm}
\normalem
\begin{algorithm}[t]
    \centering
    \resizebox{0.9\textwidth}{!}{
    \begin{minipage}{\linewidth}
    \SetAlgoLined
    \textbf{Inputs}: input audio $x$, text dict $Y$, ASR system $\mathrm{ASR(\cdot)}$, optimization numbers $max\_epoch$. \\
    \textbf{Parameters}: perturbation boundary $\epsilon$, weight coefficients in Eq. (\ref{eq_total}) $\alpha$ and $\beta$. \\
    \textbf{Output}: protected audio $x'$. \\
    \nlset{1} $\delta \leftarrow \texttt{init\_perturbation}(-\epsilon, \epsilon)$; \\
    \nlset{2} $x' \leftarrow x + \delta$; \\
    \nlset{3} \For{$j \leftarrow 1$ \KwTo $max\_epoch$}{
        \nlset{4} $\mathcal{C}_1 \leftarrow \mathcal{F} \left(\mathrm{ASR}(x'), y_t \right)$; \\
        \nlset{5}\eIf{$\texttt{Untargeted\_Sim}$}{
            \nlset{6} $y_t \leftarrow  Y_{[:|\mathrm{ASR}(x)|]}$; \\
            \nlset{7} $\mathcal{C}_2 \leftarrow \sum\limits_{i=1}^k \mathrm{CS}(E_i(x), E_i(x')) + \mathrm{CS}(M(x), M(x')) $;
        }
        {
            \nlset{8} $x_t \leftarrow \texttt{select_target_speaker}(x) $; \\
            \nlset{9} $y_t \leftarrow  \mathrm{ASR}(x_t)$; \\
            \nlset{10} $\mathcal{C}_2 \leftarrow -\,\sum\limits_{i=1}^k \mathrm{CS}(E_i(x_t), E_i(x')) - \mathrm{CS}(M(x_t), M(x')) $;
        }
        \nlset{11} $\mathcal{C}_3 \leftarrow \frac{1}{F}\sum\limits_{f=1}^F \max \left(0, p_{x' - x}(f) - \theta_x(f) \right) + ||x'-x||_2$; \\
        \nlset{12} $\mathcal{C} \leftarrow \mathcal{C}_1 + \alpha \cdot \mathcal{C}_2 + \beta \cdot \mathcal{C}_3$; \\
        \nlset{13} $\delta \leftarrow \texttt{Clamp}(-\texttt{sign}(\nabla_x \mathcal{C}), -\epsilon, \epsilon)$; \\
        \nlset{14} $x' \leftarrow x + \delta$; \\
    }
    \end{minipage}
    }
    \caption{\tool.}
    \label{algori}
\end{algorithm}

Algorithm \ref{algori} provides a detailed illustration of each step that \tool utilizes to protect audio. The input data includes the audio to be protected $x$, a long text $Y$, the target ASR system, and the optimization numbers $max\_epoch$. The output data is the protected audio $x'$. Initially, the function $\texttt{init\_perturbation}()$ is employed to randomly set the initial value of $\delta$, ensuring it stays within $[-\epsilon, \epsilon]$. Subsequently, perturbation optimization is performed for $max\_epoch$ steps. In each step, $\mathcal{C}_1$ to $\mathcal{C}_3$ are calculated separately, and their weighted sum yields the objective function value $\mathcal{C}$. For calculating $\mathcal{C}_2$, \tool offers two methods, with differing target texts for each case. If untargeted protection is applied, the target text is a segment randomly extracted from the given long text $Y$, matching the length of the original text $y \leftarrow \mathrm{ASR}(x)$. For targeted protection, the target text corresponds to the transcription of the target audio $x_t$. Using $\mathcal{C}$, gradient information can be computed to optimize $\delta$, thereby generating the protected audio $x'$.

\noindent\textbf{Regarding Perturbation Generation.} Following the classical Projected Gradient Descent (PGD)~\cite{pgd} algorithm in the adversarial attack domain, we compute the gradient of the loss function for variable $x$ to derive the perturbation: $\delta = -\texttt{sign}(\nabla_x L)$, where $\texttt{sign}(\cdot)$ denotes the sign function and $L$ represents the loss function. Subsequently, $\delta$ is projected onto the $\epsilon$-ball constraint, \textit{i.e.}, $\delta = \texttt{Clamp}(-\texttt{sign}(\nabla_x L), -\epsilon, \epsilon)$. Using $\delta$, the protected audio is updated at each step as $x' = \texttt{Clamp}(x + \delta, -1, 1)$, as shown in Algorithm \ref{algori}.

\section{Comparison with Related Work}
\begin{table*}[t]
    \centering
    \caption{The detailed comparison of related works and \tool.}
    \label{table_related-work}
    \vspace{-2mm}
    \setlength\tabcolsep{2pt}
    \resizebox{0.98\linewidth}{!}{
    \setlength{\extrarowheight}{5pt}
    \begin{threeparttable}
    \begin{tabular}{c|c|c|c|c|c|c|c|c}
    \Xhline{1pt}
    \textbf{Method} & \textbf{Target} & \textbf{Type} & \textbf{Waveform} 
    & \textbf{Phrase} & \textbf{Transferability} & \textbf{Imperceptibility} 
    & \textbf{Robustness} & \textbf{Pronunciation}\\
    \hline
    AttackVC~\cite{attackvc}  
        & \multirow{2}{*}{\makecell[c]{Voice Protection of \\ Identification }}
        & \multirow{2}{*}{AEs}
        & \XSolidBrush & \multirow{2}{*}{Inference} & \XSolidBrush 
        & $\ell_\infty$ constraint & \XSolidBrush
        & \multirow{3}{*}{\XSolidBrush}\\
    \cline{1-1} \cline{4-4} \cline{6-8}
    AntiFake~\cite{antifake}
        & & & \multirow{5}{*}{\Checkmark} & & Encoder Ensemble
        & Frequency Penalty and SNR 
        & \multirow{5}{*}{\Checkmark} & \\
    \cline{1-3} \cline{5-7}
    POP~\cite{pop}
        & \multirow{4}{*}{
            \makecell[c]{Voice Protection of \\ Synthesis Quality and \\ Identification }}
        & \multirow{2}{*}{UEs} & & \multirow{2}{*}{Fine-tuning} 
        & \multirow{2}{*}{Pivotal Objective}
        & $\ell_\infty$ constraint & & \\
    \cline{1-1} \cline{7-7}  \cline{9-9}
    SafeSpeech~\cite{safespeech}
        & & & & & & STOI and STFT loss &
        & \multirow{3}{*}{\Checkmark}\\
    \cline{1-1} \cline{3-3} \cline{5-7}
    \textbf{\makecell[c]{\tool \\ (ours)}}
        & & AEs & & \makecell[c]{E2E \\Zero-shot \& \\ Fine-tuning} 
        & \makecell[c]{Encoder Ensemble \\ with Feature Extractor}
        & Psychoacoustic Model & &\\
    \Xhline{1pt}
    \end{tabular}
    \begin{tablenotes}
    \item (1) {\textbf{Waveform}: whether the perturbation is added on original waveform.
        (2) \textbf{Transferability}: the applied approach to enhance perturbation's transferability.
        (3) \textbf{Robustness}: whether the robustness has been validated.}
    \end{tablenotes}
    \end{threeparttable}
    }
\end{table*}

To provide a clearer comparison of the distinctions and advantages between \tool and prior works, we present Table \ref{table_related-work}. This table compares aspects including algorithmic design objectives, types of data protection, whether perturbation is applied on the waveform, targeted speech synthesis types (phrases), transferability, techniques for enhancing imperceptibility, robustness verification, and consideration of pronunciation-level protection.  

\tool effectively safeguards end-to-end speech synthesis systems, covering both zero-shot and fine-tuning-based scenarios. It integrates an MFCC extractor based on an encoder ensemble to conceal speaker identity at the completed audio feature level. Specifically, \tool demonstrates strong adaptability to LLM-based speech synthesis models. Moreover, it employs a psychoacoustic model to minimize human perception of injected noise. In summary, \tool achieves a more effective, robust, and perceptually superior audio protection algorithm. 

\section{Details of Experimental Information}\label{section_details}
In this section, we illustrate the detailed information of selected synthesizers and ASR systems.

\subsection{Details of Synthesizers}\label{section_tts_models}
To provide a more comprehensive comparison of the models we adopted, we create Table \ref{table_tts_models}, which outlines the following aspects: model type, industrial origin, whether the model is LLM-based, backbone architecture, vocoder used to convert latent variables into perceptible waveforms, release time (RT), and parameter settings employed in the fine-tuning process described in Section \ref{section_exp_fine-tune}.

\begin{table}[t]
    \centering
    \caption{The detailed information and comparison of selected synthesizers.}
    \setlength\tabcolsep{2pt}
    \resizebox{\linewidth}{!}{
    \setlength{\extrarowheight}{5pt}
    \begin{threeparttable}
    \begin{tabular}{ccccccccccccc}
    \toprule[1pt]\midrule[0.5pt]
    & VITS~\cite{vits} & GSV~\cite{GPT-SoVITS} 
    & CosyVocie~\cite{cosyvoice} 
    & Llasa-1B~\cite{llasa} & Llasa-8B~\cite{llasa} 
    & StyleTTS2~\cite{styletts2} & Index-TTS~\cite{indextts} 
    & FireRedTTS-1S~\cite{fireredtts-1s}  \\
    \cline{2-9}
    {\bf Type} & fine-tuning & zero-shot & zero-shot & zero-shot & zero-shot & zero-shot & zero-shot & zero-shot \\
    {\bf Industrial?}
        & \Checkmark & \Checkmark & \Checkmark
        & \XSolidBrush & \XSolidBrush & \XSolidBrush
        & \Checkmark & \Checkmark  \\
    {\bf LLM?}
        & \XSolidBrush & \Checkmark & \Checkmark & \Checkmark & \Checkmark
        & \XSolidBrush & \Checkmark & \Checkmark  \\
    {\bf Backbone} 
        & VAE & GPT2~\cite{gpt2} & Transformer & Llama3-1B~\cite{llama3}
        & Llama3-8B~\cite{llama3} & diffusion model
        & GPT2~\cite{gpt2} & \makecell[c]{semantic LM \\ acoustic LM} \\
    {\bf Vocoder}
        & Hifi-GAN~\cite{hifigan} & Hifi-GAN & Hifi-GAN
        & \makecell[c]{HifiGAN \\ iSTFTNet~\cite{istftnet}}
        & \makecell[c]{HifiGAN \\ iSTFTNet~\cite{istftnet}}
        & Vocos~\cite{vocos} & BigVGAN2~\cite{bigvgan} & semantic decoder \\
    {\bf RT}
        & 2021 & 2024 & 2024 & 2025 & 2025 & 2023 
        & 2025 & 2025 \\
    {\bf Fine-tune}
        & \makecell[c]{Full (100 / 200)} 
        & \makecell[c]{Full (50 \& 25)}
        & \makecell[c]{Full (20)}
        & \makecell[c]{LoRA (2)} 
        & \makecell[c]{LoRA (2)} 
        & \makecell[c]{Full (50)} 
        & - & - \\
    \midrule
    & Step-Audio-TTS~\cite{step-audio} 
    & Spark-TTS~\cite{spark-tts} & XTTS-v2~\cite{xtts} 
    & FishSpeech~\cite{fishspeech} & Dia-1.6B~\cite{Dia}
    & F5-TTS~\cite{f5-tts} & E2-TTS~\cite{e2-tts} 
    & VALLE-X~\cite{vallex} \\
    \cline{2-9}
    {\bf Type}  & zero-shot & zero-shot & zero-shot & zero-shot & zero-shot & zero-shot & zero-shot & zero-shot \\
    {\bf Industrial?}
        & \Checkmark & \XSolidBrush & \Checkmark 
        & \Checkmark & \XSolidBrush & \XSolidBrush
        & \Checkmark & \Checkmark \\
    {\bf LLM?}
        & \Checkmark & \Checkmark & \Checkmark 
        & \Checkmark & \Checkmark & \XSolidBrush
        & \XSolidBrush & \Checkmark\\
    {\bf Backbone} 
        & Step-Audio~\cite{step-audio} & Qwen2.5-0.5B~\cite{qwen2.5} 
        & GPT2~\cite{gpt2} & Llama~\cite{llama}
        & Transfomer & DiT~\cite{dit} & \makecell[c]{Flow matching \\ Transformer}
        & Codec\\
    {\bf Vocoder}
        & HifiGAN & decoder & HifiGAN & Firefly-GAN~\cite{fishspeech} & DAC decoder
        & Vocos~\cite{vocos} & BigVGAN~\cite{bigvgan} & Codec decoder \\
    {\bf RT}
        & 2025 & 2025 & 2024 & 2024 
        & 2025 & 2024 & 2024 & 2023 \\
    {\bf Fine-tune}
        & - & - & - & - & - & - & - & - \\
    \midrule[0.5pt]\toprule[1pt]
    \end{tabular}
    \begin{tablenotes}
        \item (1){ \textbf{Type}: whether this model can perform zero-shot TTS.
            (2) \textbf{iSTFT}: inverse Short-Time Fourier Transform.
            (3) \textbf{LLM}: whether LLM component is employed.
            (4) \textbf{Fine-tune}: the fine-tuning type and epochs.}
    \end{tablenotes}
    \end{threeparttable}
    }
    \label{table_tts_models}
\end{table}

In terms of model types, we select VITS, a classic and backbone model requiring fine-tuning, along with other mainstream zero-shot models. Among the models, eight originate from the industry, and most (12 out of 16) are LLM-based. The LLMs utilized include Qwen2.5, Llama 3, Llama, Step-Audio, GPT2, a Transformer-based model, and a Neural Audio Codec. These language models assist the synthesizer in better learning rhythm, prosody, and semantic features. The ``Fine-tune'' column in the table indicates the implementation details used for validating the end-to-end fine-tuning scenario in Section \ref{section_exp_fine-tune}: ``Full'' denotes full-parameter training, while ``LoRA'' represents using an auxiliary Low-Rank Adaptation (LoRA) adapter to learn input features. Notably, full-parameter fine-tuning of Llasa-8B demands substantial computational resources, whereas LoRA maintains low computational resource requirements while remaining effective. The second row of numbers indicates training epochs, where ``100 / 200'' in the table represents training 100 iterations for the single-speaker dataset and 200 for the multi-speaker datasets. ``50 \& 25'' means training 50 epochs for GPT and 25 epochs for SoVITS in the GPT-SoVITS model.

\subsection{Details of ASR Systems}\label{section_asr_models}
In real-world scenarios, adversaries may employ different ASR systems to recognize textual information from audio. We briefly introduce the ASR systems considered in the experiments of Section \ref{section_exp_settings}, and in this section, we present Table \ref{table_asr_models} to provide detailed comparisons of the ASR systems used in the experiments described in the Appendix \ref{section_exp_asr}. This includes differences in the acoustic models, loss function types, and recognition performance measured by the WER metric.

\begin{table}[t]
    \centering
    \caption{The detailed information and comparison of selected ASR systems.}
    \setlength\tabcolsep{2pt}
    \resizebox{0.65\linewidth}{!}{
    \setlength{\extrarowheight}{5pt}
    \begin{threeparttable}
    \begin{tabular}{cccc}
    \toprule[1pt]\midrule[0.5pt]
    \textbf{Models} & \textbf{Acoustic Model} 
    & \textbf{Loss Type} & \textbf{WER in test-clean (\%)} \\
    \midrule
    Wav2vec2~\cite{wav2vec2} & \makecell[c]{Transformer \\ \& CNN} & CTC & 3.4 \\
    Whisper~\cite{whisper} & Transformer & Cross Entropy 
        & \makecell[c]{5.0 (base) 3.4 (small)  \\ 2.9 (medium) 2.7 (large)}\\
    Conformer~\cite{conformer} & Transformer & CTC & 3.7 \\
    CitriNet~\cite{citrinet} & CNN & CTC & 4.4 \\
    \midrule[0.5pt]\toprule[1pt]
    \end{tabular}
    \begin{tablenotes}
        \item (1){ \textbf{CNN}: convolutional neural network.
            (2) \textbf{WER in test-clean}: WER value on LibriSpeech test-clean dataset.}
    \end{tablenotes}
    \end{threeparttable}
    }
    \label{table_asr_models}
\end{table}

From Table \ref{table_asr_models}, we observe that the selected models incorporate two common backbone architectures, \textit{i.e.}, Transformer and CNN, and employ diverse loss function types. We specifically include the Whisper model~\cite{whisper}, a multilingual ASR system known for its high recognition accuracy. The largest variant, large-v3, achieves a WER of only 2.7\%, making it the best-performing model among those selected. These seven ASR systems across four categories effectively represent current mainstream technologies in the field of automatic speech recognition.

\section{Adaptive ASR Systems}\label{section_exp_asr}
In Section \ref{section_exp_fine-tune}, Section \ref{section_exp_zero-shot}, and Section \ref{section_exp_robust}, we have evaluated the protected effectiveness against the Wav2vec2 model. In the real world, adversaries can employ more types of ASR systems. Therefore, \tool should be effective when utilizing different ASR models for text recognition. In this section, we test the protective performance across six other ASR models.

\begin{table}[t]
    \centering
    \caption{The protective effectiveness and imperceptibility targeting adaptive ASR models.}
    \resizebox{0.65\linewidth}{!}{
    \begin{threeparttable}
    \begin{tabular}{ccccccc}
    \toprule[1pt]\midrule[0.5pt]
    \multirow{3}{*}{\textbf{Model}} 
    & \multicolumn{4}{c}{\textbf{Effectiveness}}
    & \multicolumn{2}{c}{\multirow{2}{*}{\textbf{Imperceptibility}}} \\
    \cmidrule(r){2-5}
    & \multicolumn{2}{c}{\textbf{VITS}}
    & \multicolumn{2}{c}{\textbf{GSV}} \\
    \cmidrule(r){2-3}\cmidrule(r){4-5}\cmidrule(r){6-7}
    & WER($\uparrow$) & SIM($\downarrow$)
    & WER($\uparrow$) & SIM($\downarrow$)
    & SNR($\uparrow$) & PESQ($\uparrow$) \\
    \midrule
    Whisper-base~\cite{whisper}
        & 99.598  & 0.144 
        & 101.082 & 0.088
        & 19.362 & 2.053 \\
    Whisper-small~\cite{whisper} 
        & 93.996 & 0.164
        & 103.518 & 0.138
        & 19.000 & 2.077 \\
    Whisper-medium~\cite{whisper}
        & 126.298 & 0.171
        & 109.666 & 0.173
        & 19.236 & 2.089 \\
    Whisper-large-v3~\cite{whisper}
        & 84.618 & 0.163
        & 66.534 & 0.164
        & 19.245 & 2.019 \\
    Conformer~\cite{conformer} 
        & 105.145 & 0.126
        & 81.753 & 0.180
        & 12.835 & 1.638 \\
    CitriNet~\cite{citrinet} 
        & 89.520 & 0.137
        & 59.921 & 0.302
        & 12.948 & 1.629 \\
    \midrule[0.5pt]\toprule[1pt]
    \end{tabular}
    \end{threeparttable}
    }
    \label{table_exp_asr}
\end{table}

Table \ref{table_exp_asr} shows the protection effectiveness and imperceptibility of \tool across different ASR models. The results demonstrate that our method provides strong protection for various ASR models, with WER values consistently above 50\%, indicating low intelligibility of synthesized audio, and SIM values below the threshold of 0.25~\cite{ecapa_tdnn}, indicating low similarity to the original audio. Whisper-large-v3, known for its superior text recognition accuracy, ease of utilization, and multilingual capabilities, is widely adopted in the industry. \tool also effectively protects against Whisper-large-v3, achieving average WER and SIM values of 75.576\% and 0.163, respectively, across two models. Additionally, the imperceptibility of perturbations generated for different models outperforms strong baselines \textit{i.e.}, AntiFake, POP+ESP, and SafeSpeech.

\section{Multilingual and Multi-Speaker Evaluation} \label{section_exp_multi}
In Section \ref{section_exp_fine-tune} and Appendix \ref{section_exp_asr}, we evaluate the effectiveness and transferability of the proposed \tool on a single-speaker English dataset, LibriTTS. However, adversaries may encounter diverse speech samples, including multi-speaker and multilingual scenarios. To address this, we further validate our method on CMU ARCTIC, a multi-speaker dataset, and THCHS30, a multi-speaker Mandarin dataset targeting the Wav2vec2 model. Specifically, we fine-tune VITS and GSV models on CMU ARCTIC while using the GSV model for fine-tuning on THCHS30.

\begin{figure}[t]
    \centering
    \begin{subfigure}{0.35\textwidth}
        \centering
        \includegraphics[width=\textwidth]{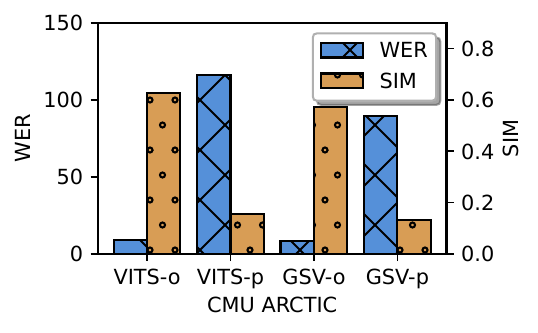} 
        \caption{Multi-speaker dataset.}
        \label{fig_multi_speaker}
    \end{subfigure}
    \begin{subfigure}{0.25\textwidth}
        \centering
        \includegraphics[width=\textwidth]{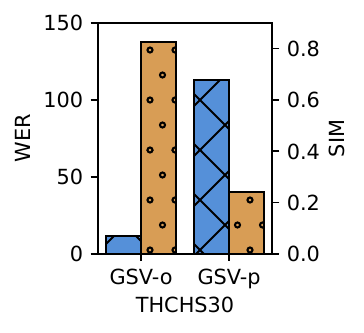}
        \caption{Mandarin dataset.}
        \label{fig_multilingual}
    \end{subfigure}
    \vspace{-0.5em}
    \caption{Test results of Multi-speaker and Mandarin datasets.}
    \label{fig_multi}
\end{figure}

Figure \ref{fig_multi_speaker} illustrates the experimental results on the multi-speaker dataset (``-o'' represents fine-tuning on the original dataset, and ``-p'' denotes fine-tuning on the protected dataset in the figure). We can find that both models are capable of synthesizing high-quality audio with corresponding speaker timbres on clean samples. After fine-tuning the audio protected by \tool, the WER and SIM averaged 103.021 and 0.144, respectively, indicating that \tool can effectively prevent the pronunciation and timbre information in a multi-speaker end-to-end fine-tuning scenario. Figure \ref{fig_multilingual} demonstrates the fine-tuning effect of GSV on the THCHS30 dataset, where the recognition ASR system uses a multilingual recognition model, whisper-base, as the target for adversarial attacks.

\section{Human Study}\label{section_exp_human}
In our previous experiments, we have validated the effectiveness of \tool and its perception through objective evaluation metrics. However, we also need to verify the subjective perception of audio by human ears, as synthesized audio in the real world needs to interact with humans. Therefore, this experiment conducts a subjective evaluation to validate human perception and discrimination of synthesized audio, as well as the perception of protected audio.

\begin{table}[t]
    \centering
    \begin{minipage}{0.4\textwidth}
        \centering
        \caption{The subjective evaluation of the ground truth (GT) and \tool-protected dataset.}
        \label{table_human_real}
        \setlength\tabcolsep{2pt}
        \resizebox{0.8\textwidth}{!}{
        \begin{tabular}{cc}
        \toprule[1pt]\midrule[0.5pt]
            & {\bf MOS}($\uparrow$) \\
            \midrule
            GT & 4.788 $\pm$ 0.157 \\
            \tool (UT) & 3.522 $\pm$ 0.218 \\
        \midrule[0.5pt]\toprule[1pt]
        \end{tabular}
        }
    \end{minipage}
    \quad
    \begin{minipage}{0.5\textwidth}
        \centering
        \caption{{Human perceptual evaluation of the quality and intelligibility of the synthesized speeches by different training samples.}}
        \label{table_human_sythetic}
        \setlength\tabcolsep{2pt}
        \resizebox{0.9\textwidth}{!}{
        \begin{tabular}{ccc}
        \toprule[1pt]\midrule[0.5pt]
            & {\bf MOS}($\downarrow$) & {\bf Intelligibility}($\downarrow$) \\
            \midrule
            clean & 4.842 $\pm$ 0.123 & 100.000\% \\
            AntiFake & \uline{2.055 $\pm$ 0.309} & 99.074\% \\
            POP+ESP & {\bf 0.851 $\pm$ 0.364} & 89.814\% \\
            \midrule
            {\bf \tool (UT)} & 2.615 $\pm$ 0.331 & \uline{50.925\%} \\
            {\bf \tool (T)} & 3.060 $\pm$ 0.373 & {\bf 8.333\%} \\
        \midrule[0.5pt]\toprule[1pt]
        \end{tabular}
        }
    \end{minipage}
\end{table}

\noindent\textbf{Recruitment Process.} This subjective survey has been approved by the Human Ethics Research Committee at the first author's institution. We create the questionnaire through Credamo and recruit 36 volunteers to participate in the survey, which is comparable to similar studies, such as AntiFake of 24 participants. All volunteers are over 18 years old and possess good English skills. Their average response time is 200.194 seconds.

\noindent\textbf{Filtering.} We prohibit volunteers under 18 from participating in the questionnaire. Within the questionnaire, we include two simple random arithmetic questions, and incorrect answers result in rejection. We also filter out all non-serious responses, \textit{e.g.}, the same answers across all questions, or excessively short response times.

\noindent\textbf{Questionnaire Setup.} We establish two sections for subjective testing for the synthesized audio and protected audio, with a total of 22 samples.

\noindent\textbf{Task 1: Study on Protected Speech.} In the subjective test of protected audio, we select 3 audio samples protected by \tool to test naturalness and similarity to the original audio. In order to improve the confidence level of the subjective experiment and reduce the potential bias, we calculate the MOS by taking into account the 95\% confidence intervals, which can be found in the previous research~\cite{safespeech}. Results in Table \ref{table_human_real} show that the MOS of 3.522$\pm$0.218 suggests that the embedded perturbations do not significantly reduce normal audio usability, and the distortions are acceptable to human ears when the MOS value surpasses 3.0~\cite{antifake}.

\noindent\textbf{Task 2: Study on Synthetic Speech.} In this part, we select 3 synthesized audio samples trained on original samples, baseline-protected methods, and \tool-protected samples. Table \ref{table_human_sythetic} presents the experimental results for Task 2, revealing that compared to synthesis from the original audio.
For synthetic audio evaluation, we utilize audio quality (MOS) and pronunciation intelligibility. Audio quality assesses noise levels and perceptual quality, calculated consistently with Task 1. The ESP method exhibits the worst synthesis quality because its perturbation addition process causes the most severe distortion to the original audio, rendering it unusable. Pronunciation intelligibility involves presenting participants with both audio and its corresponding prompt text to judge whether the audio content matches the given text. The table results show that \tool effectively disrupts the model's original pronunciation patterns: only 50.925\% (UT) and 8.333\% (T) of participants perceived correct pronunciation alignment with the text, while most participants identified mismatches, demonstrating significant improvement over previous baselines.

Through human study, we observe two key findings: (1) Human auditory perception aligns with objective metrics from prior experiments; (2) Experimental results confirm that \tool's noise injection not only bypasses human auditory detection but also substantially reduces the probability of participants being deceived by deepfake audio.

\end{document}